\documentclass[11pt]{article}

\usepackage{graphicx}

\usepackage{times}
\usepackage{listings} 
\usepackage{amssymb}  

\usepackage{algorithm}
\usepackage{algorithmicx}

\begin{document}

\title{RWN:  A Novel Neighborhood-Based Method for
Statistical Disclosure Control}
\author{
    Noah Perry, Department of Statistics, University of California, Davis \\
    Norman Matloff, Department of Computer Science, University of California, Davis \\
    Patrick Tendick, FINRA}

\maketitle

\begin{abstract}

A novel variation of the data swapping approach to statistical
disclosure control is presented, aimed particularly at preservation of
multivariate relations in the original dataset. A theorem is proved in
support of the method, and extensive empirical investigation is
reported.

\end{abstract}

\section{Introduction}   

The field of statistical disclosure control (SDC)---maintaining privacy
of individual records in a dataset while retaining the statistical
utility of the data---has long been the subject of arcane technical
analysis, conducted mainly by statisticians.  The advent of
\textit{differential privacy} (DP) methods in 2006 \cite{dwork2006}
brought computer scientists into the field, and the arcane nature of the
field changed with the well-publicized adoption of DP by the United
States Census Bureau in 2017.  That move by the Bureau has become
somewhat controversial \cite{wp} \cite{sarathy} and though we do not
address that controversy, we note the salutary effect of greatly
increasing public awareness of SDC.

Different SDC tools may be appropriate for different databases in
different settings, not just in terms of numeric degree of protection
afforded by a tool, but also in terms of usability, interpretatibility
and transparency, for end users \cite{oberski}.  Here we
develop new methodology that we believe database administrators (DBAs)
will find useful in a variety of settings.  

Our proposed method, Randomization within Neighborhoods (RWN), to be
presented below, is inspired by data swapping, a classic approach to
SDC.  However, RWN differs from previous data swapping approaches, in
that it exploits a certain statistical independence property, to be
described in detail in the next section. 

This paper is organized as follows.  We give a brief overview of SDC
methods in Section \ref{sdcmethods}, followed by a discussion of
considerations in SDC specific to databases used for statistical
analysis in Sections \ref{statviews} and \ref{daypass}.  The RWN method
is presented in Section \ref{rwn}.  The underlying theory is given in
Section \ref{theory}. Tuning parameter selection is 
covered in Section \ref{tune}.  Our empirical investigation is
discussed in Section \ref{empirical}, and computational issues are discussed
in Section \ref{large}.

Note by the way that while we refer for convenience at some points to
census data as a concerete example, our methodology is meant to be
general, not specific to the census.  It could be applied to medical
data, employee data and so on.

\section{SDC Methods}
\label{sdcmethods}

Good surveys of SDC methods are in \cite{duncan} \cite{hundepool}
\cite{brand}.  In order to contrast with RWN, we give a brief overview
here. 


Previously the Census Bureau had used the popular \textit{data swapping}
method for SDC \cite{fischetti} \cite{duncan} \cite{mcintyre}.  A set of
\textit{key variables} is defined that may render certain individual
records in the data vulnerable to disclosure.  Some records, especially
those deemed most at risk to disclosure, will have the values of their
key variables swapped with those in other records, say drawn from the
same geographic region.

There are variants, notably \textit{data shuffling} \cite{shuffle}.
These methods are typically applied one variable at a time, thus
creating the concern of attenuation of multivariate relations.

The Bureau has also used \textit{cell suppression}, in which any query
concerning a very small number of database records is denied.
However, in 2017, the Bureau, after performing various simulations,
decided that the swapping approach was in danger of reconstruction
attacks and turned to DP.

Another standard SDC approach is \textit{data perturbation}, in which
random noise is added to achieve privacy.  The Census Bureau has used
this method in the past as well, 

DP is also a perturbation method. Its novelty, though, is in its ability
to be able to quantify the degree of privacy, in a manner having certain
mathematical traits, such as composability.  Another difference from
classic noise addition methods is that in DP, the noise is typically
added to a statistic of interest, say a mean (\textit{global} DP),
rather than to the microdata itself (\textit{local} DP).

Two of the present authors, NM and PT, have a data perturbation
background background \cite{matloffoak} \cite{matloffifip}
\cite{tendicktods} \cite{tendickplaninf}, and were interested in
modernizing that approach, making a proposal in 2016 \cite{mtarxiv} for
a novel SDC method inspired by data swapping.  The present work, joined
by author NP, develops those ideas.  

\section{Statistical Views and Goals}
\label{statviews}

Leo Breiman's famous essay \cite{breiman} on predictive modeling
described a ``cultural'' difference between researchers in statistics
and computer science, the former viewing the world more in terms of
probabilistic behavior, the latter in terms of algorithms.  Curiously,
as pointed out by statistician Larry Wasserman \cite{wass}, a similar
difference later arose in the SDC field.  Referring to the need in
(global) DP to develop a separate DP-compliant method for each
statistical procedure used (mean, regression analysis etc.), he noted
the contrasting views: 

\begin{quote}

CS view:  Receive a query for a [specific statistical procedure], return
a private answer.

Statistics view:  Give me data. Then I can: draw plots, fit
models, test fit, estimate parameters, make predictions ...

\end{quote}

An amusement park metaphor will be useful.  Under the CS approach, one
must buy a separate ticket for each ride.  Some rides won't be available
at all, pending development of tickets tailored to those rides.  With
statistics, one purchases a day pass, good for all rides.  Our focus
here is on settings in which one wishes to have a ``statistical day
pass.''  We protect the data in some way, say perturbation, then let
users conduct whatever types of statistical analyses they wish in an
open-ended way.

Good arguments can be made for either view.  As noted, the
one-query-at-a-time nature of the DP/CS approach enables the setting of
precise guarantees of privacy, which would be difficult or impossible in
the statistics approach.  On the other hand, this means the user is
restricted to only the types of queries for which a DP version has
already been developed and implemented.  Linear regression may be
available, say, but not quantile regression.

(We note in passing that unlike most applications of DP, the US Census
Bureau's DP methodology does amount to a ``day pass.'' This is because they
add noise to the raw data, which are cell counts in a huge contingency
table, rather than to the output of a statistical analysis, such as a
mean.)

Due to our goal of developing a statistical \textit{analysis} ``day
pass,'' in the sense Wasserman described---estimating parameters,
fitting models and so on, open-endedly---we look at the data in the usual
\textit{statistical} manner, i.e.\ as a sample from a population.  This
is in contrast to many SDC applications in which the data themselves are
of primary interest, say the total count of people in a given income
range for a given census block.

A typical example might be that of a medical database, in which the
privacy of individual patients is required, but with which medical
researchers can still conduct statistical analyses, making population
inferences.

We suppose here that there is some population value $\theta$ for which
we wish to obtain a sample estimate $\widehat{\theta}$, performing
statistical operations such as inference (confidence intervals,
hypothesis tests).  These operations will be conducted on the perturbed
data obtained by applying RWN to our original microdata.  Typically
$\theta$ will be vector valued, such as a vector of regression
coefficients.

\section{Desirable Statistical ``Day Pass'' Characteristics}
\label{daypass}

In developing a new SDC procedure, such as our proposed RWN, 
these goals are key:

\begin{itemize}

\item Ability to handle mixed continuous and discrete/categorical data.

\item Preservation, to the degree possible, of not only univariate but
also multivariate distributions/relations.

\item Limiting the increase in size of the standard errors of
$\widehat{\theta}$.

\item Preservation, to the degree possible of statistical inference
levels related to $\widehat{\theta}$.

\end{itemize} 

\noindent 
On the other hand, as noted, we do not take as a goal the preservation
of marginal totals as in Census data.

Let's elaborate a bit on these goals.

\subsection{Handling Mixed Continuous and Categorical Data}

A major obstacle to data perturbation methods is their inability to
handle discrete/categorical data.  Consider a variable such as Number of
Children in Family.  After noise addition, a value may become negative,
an unacceptable situation.  A similar difficulty arises with categorical
variables, after they are converted to dummy (\textit{one-hot}) form. 

Indeed, the vast majority of the Census Bureau's TopDown algorithm
\cite{topdown} is devoted to making adjustments to negative values, and
satisfying certain constraints involving marginal totals.

RWN will be seen to handle mixed continuous and discrete/categorical data
in a simple, natural manner.

\subsection{Preservation of Multivariate Relations}

Absent some compensating feature, any change to the data arising from
applying a ``day pass'' SDC procedure, say perturbation or swapping,
will result in distortions of the relations between variables
in the data.  This will also occur with cell suppression methods.  Since
analysis of multivariate relations comprise the very core of statistics, we take
as a major goal at least approximately preserving such relations. 

We are of course willing to let those relations be one aspect of the
utility/privacy tradeoff that is necessary to any disclosure avoidance
technique.  Let's call this property Multivariate Relations Attenuation
Resistance (MRAR).  

The goal then is to develop an SDC method that includes MRAR, with the
method providing the user a ``lever'' that she can use to choose
her desired utility/privacy tradeoff level.

Comparatively little work in the SDC field has focused on MRAR.  It is
mentioned only briefly in \cite{hundepool} and \cite{duncan} --- and no
wonder, as MRAR is a challenging condition to meet.  

Consider noise addition methods.  One actually can preserve second-order
moment structure by setting the covariance matrix of the noise to that of
the data \cite{kim} \cite{matloffoak} \cite{tendickplaninf}.  But higher-order
moments are lost and other distortions can occur.  And there are no
obvious techniques for extending this property with noise addition in
mixed continuous/categorical variable settings.  

\section{RWN:  Randomization within Neighborhoods}
\label{rwn}

The method works roughly as follows. For each record in the data, we
define a neighborhood using either a Euclidean distance-based radius or
k-nearest neighbors.  Then, for each record $r$ we randomly choose a subset
of the variables to perturb.  For each such variable, we replace its
original value by its counterpart in a randomly chosen record in the
neighborhood of $r$.  A key point is that a different random neighbor
record is used for each of the variables to be perturbed in $r$.

More formally:   

Let $W = (w_{ij}), i = 1,..., n, j = 1,...,p$ denote our original data
on $n$ individuals and $p$ variables and $W' = (w'_{ij}), i = 1, ..., n,
j = 1, ..., p$ be the released (i.e.\ perturbed) data.

Choose neighborhood radius $\epsilon > 0$, or number of nearest neighbors $k$,
and modification probability $q$.  Then we form our released data
$W'$ as follows:

\begin{quote}

For $i = 1,...n$:

\begin{enumerate}

\item Consider record $i$ in the database:

\begin{equation}
r_i = (w_{i1},...,w_{ip})
\end{equation}

\item Find the set $S_i$ of records within the neighborhood 
of $r_i$ other than $r_i$ itself.  Each neighborhood is defined to be
either the $k$-nearest neighbors of $r_i$ or the set of neighbors 
within $\epsilon$ distance of $r_i$, whichever set is larger.


\item For $j = 1,...,p$:


        With probability $1-q$, leave 
        $w_{ij}$ unmodified, but with probability $q$, modify it. For
        variables that are chosen to be modified, we replace $w_{ij}$
        with the value in variable $j$ of a random record in the
        neighborhood $S_i$.  As noted, there will be a different such
        random record for each $j$.
        This results in a perturbed data point
        $w'_{ij}$. For unmodified variables, $w'_{ij} = w_{ij}$.


\item Store the released, modified version of $r_i$ as

\begin{equation}
r_i' = 
(w_{i1}',...,w_{ip}')
\end{equation}

\end{enumerate}

\end{quote}

A key point is that in Step 3, the $p$ actions here are \textit{taken
independently of each other}.  In other words, the process acts as if the
$p$ variables in the data are statistically independent of each other.
This would at first seem to violate our goal of MRAR, but it is all
resolved in the theorem in Section \ref{theory} below.

We call this technique Randomization Within Neighborhoods (RWN).

\subsection{Comparison to Rank Swapping}
\label{compare}

\textit{Rank swapping} \cite{moore} does data swapping using ranks
rather than data values, in order to facilitate dealing with discrete
variables.  It's implemented as the function \textbf{rankSwap()} in the
popular R package for SDC, \textbf{sdcMicro} \cite{templ}.  It too is
neighborhood-based.  RWN differs from rank swapping in several respects:

\begin{itemize}

\item Rank swapping's MRAR feature is focused on bivarate relations,
while RWN is fully multivariate, exploiting the theorem in Section
\ref{theory}.

\item RWN's neighborhoods are more complex than rank swap's swapping ranges due the following features:
	\begin{itemize}
		\item The size of the RWN neighborhoods can differ for each record.
		\item The distances are computed based on all variables in a record instead of one variable at a time, so the records within a given neighborhood can be "similar" in many different ways.
		\item RWN duplicates data values, rather than actually swapping them. This actually helps reduce \textit{k-anonymity} issues \cite{templ}. 
 		\item Setting $q < 1$ allows for the possibility that some values may be the original ones.
	\end{itemize}

\end{itemize} 

\section{Rationale and Theoretical Basis} 
\label{theory}

Since we are using neighborhoods, one might ask, ``Why not just replace
entire data rows---a given row is replaced by a neighboring row---rather
than do replacement component by component, taking different components
from different neighboring rows?''  That would achieve our MRAR goal,
but at the expense to too much increase in $Var(\widehat{\theta})$.

By independently perturbing each element of a row, we can limit the
increase in standard errors of $\widehat{\theta}$.  However, we then
must ask whether our method has the all-important MRAR property.
The following theorem shows that it does.

\begin{quote}

Let $f_X$ be a density function for the $p$-variate vector $X =
(X_1,...,X_p)$.  Consider the conditional distribution of $X$, given
that $X$ is in a small neighborhood of $t$.  Then the $X_i$ are
approximately independent in this distribution.

\end{quote}

For expositional convenience, the theorem and proof will be stated for
the case $p = 2$.

\label{thm1}

{\bf Theorem:}  Consider a bivariate random vector $(X,Y)$ having a
joint density, and set $\epsilon > 0$.  For any $t$ in $\mathcal{R}^2$,
let $A_{t,\epsilon}$ denote the $\epsilon$ neighborhood of $t$.  Let $F$
denote the joint cdf of $(X,Y)$.  Given $(X,Y) = t$, define
$G_{t,\epsilon}$ to be the cdf of $(X,Y)$, given that that vector is in
$A_{t,\epsilon}$.  Finally, given $(X,Y) = t$, define
\underline{independent} random variables $U$ and $V$ to be drawn
randomly from the first- and second-coordinate marginal distributions of
$G_{t,\epsilon}$, respectively.  Then

\begin{equation}
\label{thetheorem}
\lim_{\epsilon \rightarrow 0}
P
\left (
U \leq a \textrm{ and } V \leq b
\right )
= F(a,b)
\end{equation}

for all $-\infty < a,b < \infty$.  

In other words, as $\epsilon$ goes to 0, the bivariate
distribution of $(U,V)$ goes to that of $(X,Y)$, {\it even though $U$
and $V$ are independent while $X$ and $Y$ are not independent}.

$\square$

\bigskip

Note that (\ref{thetheorem}) concerns the unconditional distribution of
$(U,V)$.  The latter is a random vector in $A_{(X,Y),\epsilon}$.

\bigskip


{\bf Proof:}

First,

\begin{equation}
\lim_{\epsilon \rightarrow 0} U = X
\end{equation}

and

\begin{equation}
\lim_{\epsilon \rightarrow 0} V = Y
\end{equation}

Using the Bounded Convergence Theorem, we have  


\begin{eqnarray}
\lim_{\epsilon \rightarrow 0}
P
\left (
U \leq a \textrm{ and } V \leq b
\right )
&=& 
\lim_{\epsilon \rightarrow 0}
E \left [
P
\left (
U \leq a \textrm{ and } V \leq b
~|~ X,Y \right )
\right ] \\ 
&=& 
\lim_{\epsilon \rightarrow 0}
E \left [
P(U \leq a ~|~ X,Y ) \cdot
P(V \leq b ~|~ X,Y ) 
\right ] \\
&=& E \left [
1_{X \leq a} \cdot
1_{Y \leq b}
\right ] \\
&=& E \left [
1_{X \leq a \textrm{ and } Y \leq b}
\right ] \\
&=& P(X \leq a \textrm{ and } Y \leq b) \\
&=& F(a,b)
\end{eqnarray}


\section{Neighborhoods and Tuning Parameters}
\label{tune}

The neighborhoods are formed using both the Euclidean
distance-based radius $\epsilon$ and the number of nearest neighbors $k$,
which must be specified by the user.  In short, $\epsilon$ provides control
over the similarity of the data points within a neighborhood to each
other while the nearest neighbor parameter $k$ controls the minimum size
for the neighborhood.

For many datasets, there are typical records as well as records with
more unusual or extreme values. Typical records will have many neighbors
even for small values of $\epsilon$ while unusual records may have zero
neighbors unless $\epsilon$ is large.  If $\epsilon$ alone were used to form
neighborhoods, RWN would set these unusual records with empty
neighborhoods to contain missing values in all variables. This would
protect their privacy, which is important considering that these unusual
records may correspond to more identifiable individuals, but would
result in a complete loss in their utility. 

\begin{figure}[tb]
    \centerline{
    \includegraphics[width=5.0in]{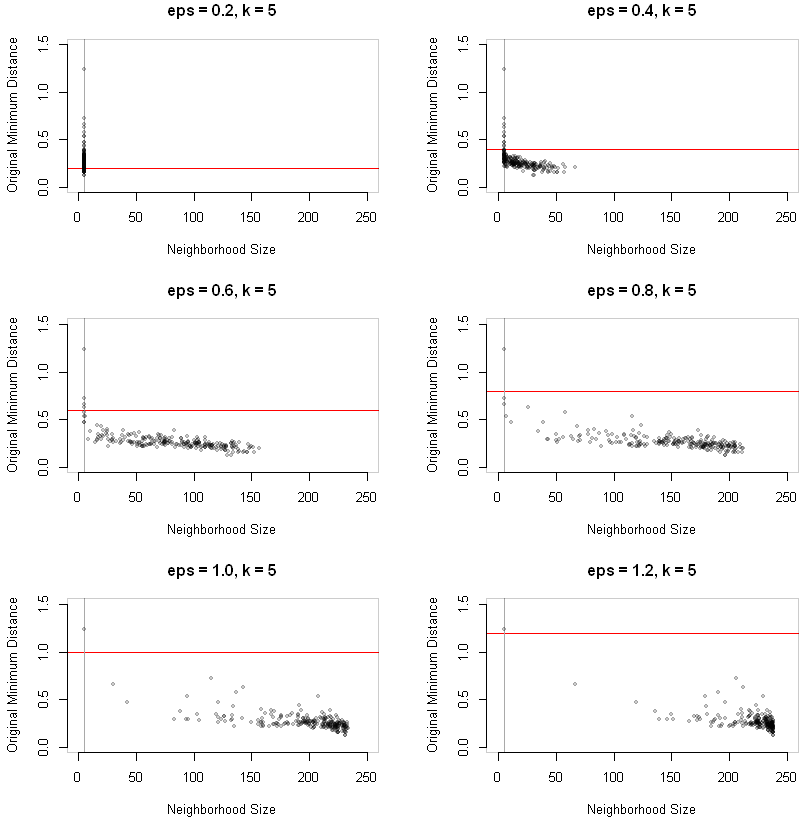}
    }
    \caption{Body Fat Data, Plots of Minimum Distance vs. Neighborhood Size}
    \label{fig:bf_mindist_neighborhood}
\end{figure}

To avoid this, the user may
choose to increase $\epsilon$ until these more extreme records have non-empty
neighborhoods. However, as seen in Figure \ref{fig:bf_mindist_neighborhood}, an increase in
$\epsilon$ can substantially increase the size of the neighborhood for
typical records as well, causing the values in a typical record to be
mixed with very dissimilar records in the perturbation process,
potentially leading to a decrease in utility of the perturbed data.  

On the other hand, using $k$ alone would impose both an upper and lower
bound on the neighborhood size. For instance, for small $k$, the
neighborhood size may be suitable for unusual records but unnecessarily
small for typical records.  

Thus, having two neighborhood size
parameters $\epsilon$ and $k$ gives the Data Stewardship Organization (DSO)
finer control over the perturbation of the data to balance utility with
privacy for a specific dataset.


In Figure \ref{fig:bf_mindist_neighborhood} we illustrate how RWN
interacts with a specific dataset and choice of tuning parameters to
form neighborhoods for perturbation.  Using the \textbf{bodyfat} data,
we calculate the minimum distance for each record and plot it alongside
the neighborhood size (i.e. the number of records within the
neighborhood) for multiple choices of $\epsilon$ while holding constant
$q = 1$ and $k = 5$.  In the charts, the red horizontal line depicts the
value of $\epsilon$ and the gray vertical line depicts the value of $k$. 

\section{Empirical Investigations}
\label{empirical}

We will present the results of our empirical investigation shortly.  But
first, we describe our criteria.

Any SDC method is a balance of statistical utility and degree of
privacy.  Empirical investigation of the method must then define
measures for these two criteria.   

\subsection{Statistical Utility}
\label{util}

As noted, statistical analysis is at its core a matter of identifying
relations between variables.  The question to consider in the SDC
context is whether the relationships that exist in the original data
tend to remain intact in the released data.  Here we follow \cite{king},
who note that a reasonable measure is to assess whether the released
dataset ``can obtain approximately the same substantive [relational]
results while simultaneously protecting the privacy.''  This is one of
the approaches we take below, using color correlation plots such as in
Figure \ref{fig:bf_corrplot1}.  We find that we can perturb data while
retaining broad correlation structure, including to a large extent the
strength of the correlations.

Another aspect of utility is validity of standard errors for statistical
inference purposes.  For SDC methods affecting only a small portion of
the data, this is less of an issue.  For RWN, one can prove, say, that
the standard errors are asymptotically valid, as $q \rightarrow 0$, and
find this to hold in our empirical work below.  (See \cite{king} for a
DP solution, under certain assumptions.  As usual, the problem of
discrete/categorical variables remains a challenge.)

Since we emphasize MRAR, a utility measure is needed toward that end.
We address this by investigating how correlations, regression coefficients,
and principal components are affected by perturbation.

Most SDC methodology is aimed at estimating relationships, rather than
using those relationships for prediction of new cases.  Here, we
investigate that latter aspect.  

\subsection{Privacy}

A number of measures of privacy have been used in the SDC literature.
The reader is referred to the references on this, but here our choices
were guided primarily by two examples.  (As already mentioned, RWN by its
nature is resistant to k-anonymity problems.)

\begin{itemize}

\item [(a)] The \textbf{bodyfat} dataset (see later section for details)
has several outliers.  If an intruder to the database knows, for
instance, that a certain individual has the highest Body Mass Index of
all the subjects in the data, then this individual is at risk of the
intruder identifying sensitive variables (say if this dataset contained
private health records).

\item  [(b)]Problems may arise involving ``inliers.''  A hypothetical
example that has been used in the SDC literature is that of an employee
database, in which an intruder knows that there is just one female
electrical engineer.  The intruder then queries the total salaries of
all female electrical engineers, and thereby illicitly learns her
salary.

\end{itemize} 

A number of measures could be used to gauge potential problems of these
sorts.  We address outlier issues such as in Example (a) by using
Mahalanobis distance, and via Cook's distance for linear regression
analysis.  We measure inlier problems via distance to closest neighbor.

\subsection{Experiments}

The first dataset used for the empirical experiments is the
\textbf{bodyfat} dataset from the \textbf{mfp} R package \cite{mfp}. The
raw dataset contains body measurements of 252 adult males and two
estimates of their body fat percentage calculated using the Brozek and
Siri equations.  For our analysis, we use only the body fat percentage
based on the Siri equation, eliminate 11 observations that contain
values that appear to be erroneous or biologically implausible such as
body fat percentages less than four percent, and calculate Body Mass
Index (BMI) for each individual.

For the experiments in this section, we use the cleaned \textbf{bodyfat} 
dataset as well as several perturbed versions of this dataset created 
using RWN with tuning parameters $k = 5$, $q = 1$, 
and varying values of $\epsilon$. In each figure where a horizontal blue line 
is present, it denotes the result for the unperturbed data.



\textit{Correlation:}

\begin{figure}[tb]
\centerline{
\includegraphics[width=5.3in]{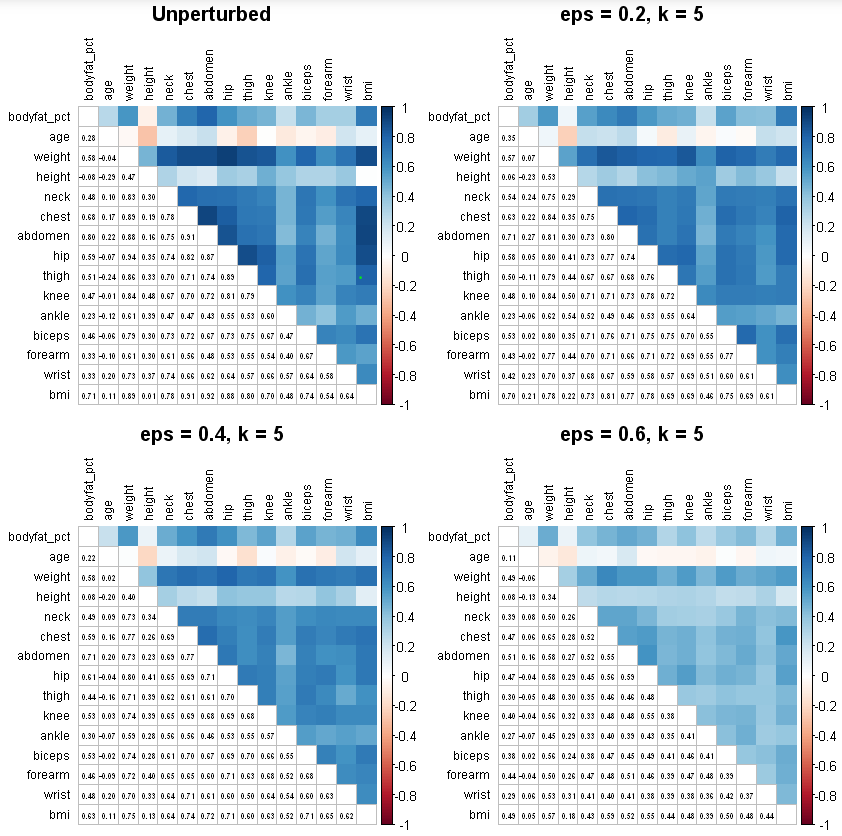}
}
\caption{Body Fat Data, Correlation Plot}
\label{fig:bf_corrplot1}
\end{figure}

We calculated Pearson correlation coefficients pairwise for all
variables in the \textbf{bodyfat} dataset.  The color correlation plots
in Figure \ref{fig:bf_corrplot1}
show that for small $\epsilon$, the correlation coefficients in the
perturbed and unperturbed data for a given variable are of the same sign
and of very similar magnitude. This provides evidence of the MRAR
property of RWN. 

\begin{figure}[tb]
\centerline{
\includegraphics[width=5.3in]{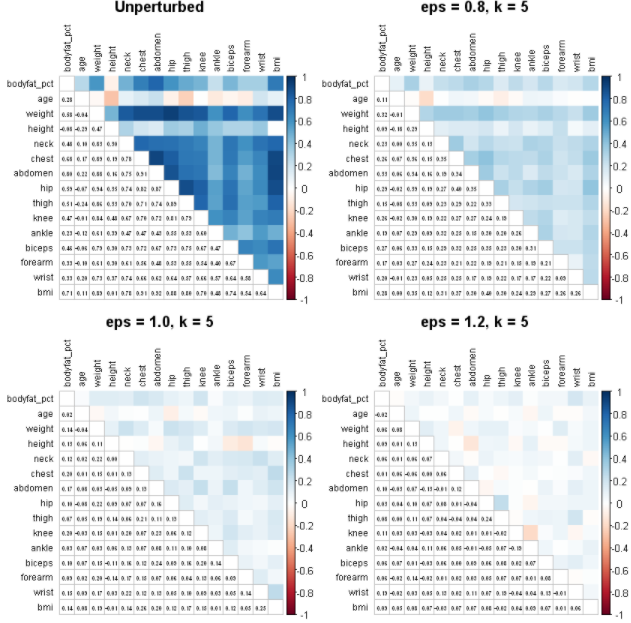}
}
\caption{Body Fat Data, Correlation Plot (Large eps)}
\label{fig:bf_corrplot2}
\end{figure}    


\textit{Regression:}

We regressed the body fat percentage on BMI and the neck, chest,
abdomen, and hip measurements.  After estimating the model on each datasets, we
compared the estimated coefficients, standard errors, and Cook's distances.  

\begin{figure}[tb]
    \centerline{
     \includegraphics[width=5.0in]{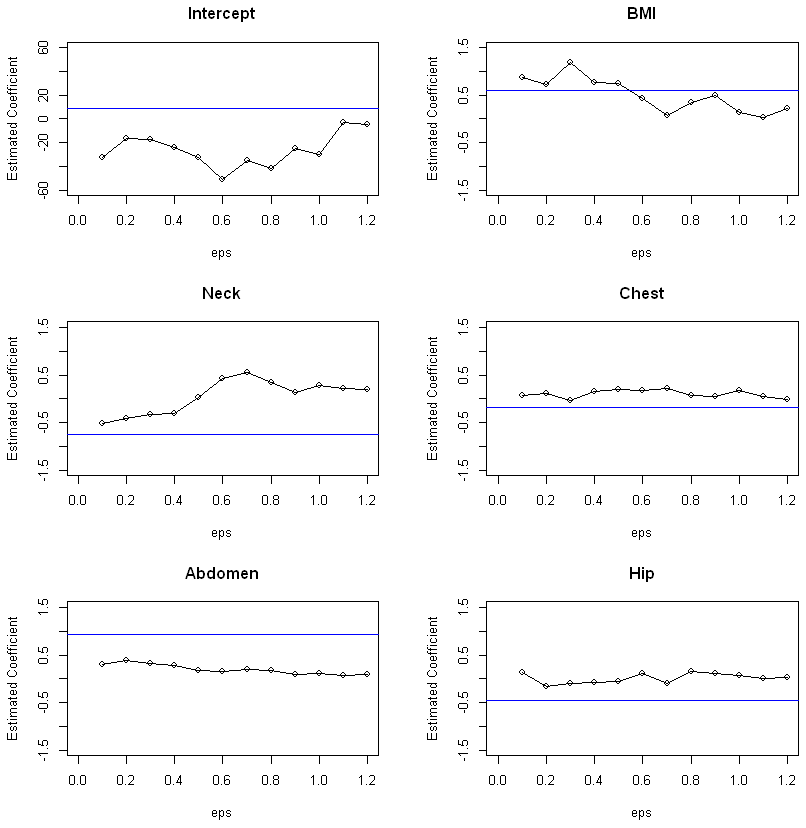}
    }
    \caption{Body Fat Data, Regression Coefficients}
    \label{fig:bf_reg_coefs}
\end{figure}

As seen in Figure \ref{fig:bf_reg_coefs}, there is
non-trivial variation in the estimated coefficients on the intercept,
BMI, and neck variables over varying $\epsilon$.  We also note that
there are numerous cases where the sign of the estimated coefficient on
the perturbed data differs from the sign of the corresponding estimated
coefficient in the model estimated on the unperturbed data.  To some
extent, these results are unsurprising.  It is clear from the
correlation plot that many of the body measurements have a high
positive correlation with each other. Consequently, we expect some
instability in the estimated coefficients due to multicollinearity. 

\begin{figure}[tb]
    \centerline{
     \includegraphics[width=5.0in]{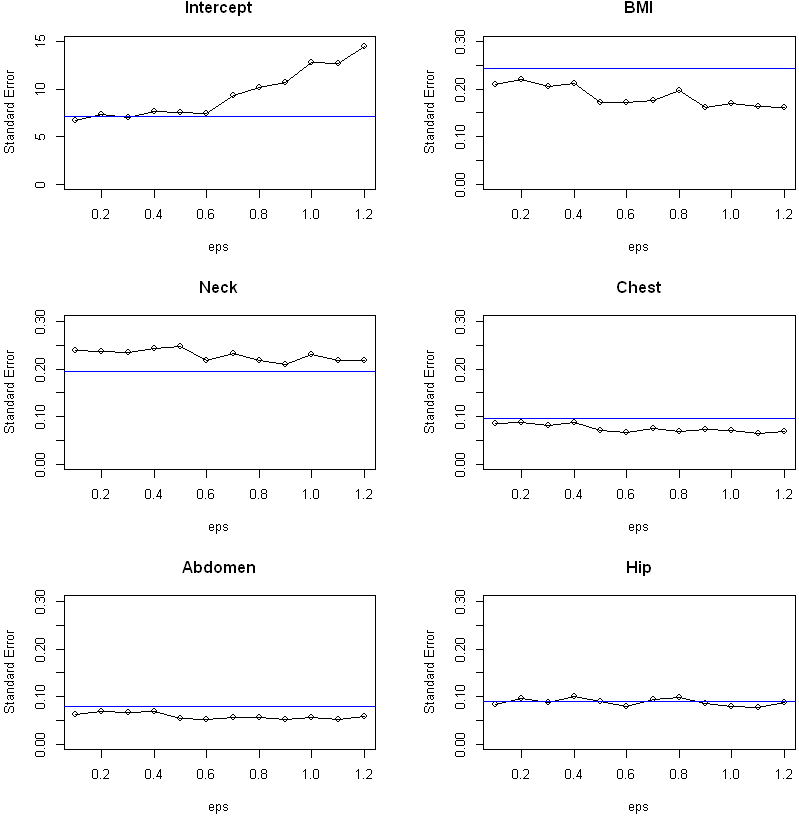}
    }
    \caption{Body Fat Data, Regression Standard Errors}
    \label{fig:bf_reg_se}
\end{figure}

As shown in Figure \ref{fig:bf_reg_se}, with the exception
of the intercept for larger $\epsilon$ values, 
the standard errors in the perturbed data appear to be 
relatively stable and similar to the standard errors 
calculated using the unperturbed data.

\begin{figure}[tb] \centerline{
    \includegraphics[width=5.0in]{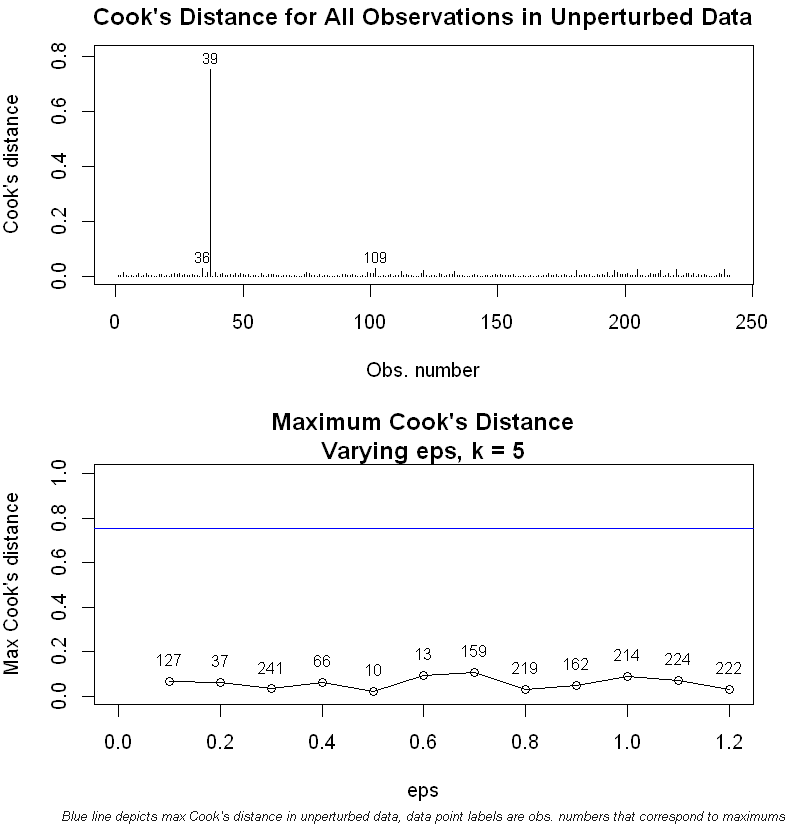} }
    \caption{Body Fat Data, Cook's Distances} \label{fig:bf_cooks_dist}
    \end{figure}

The cleaned bodyfat dataset contains one individual who is substantially
larger than all the others. Consequently, this individual would be one
of the most easily identifiable in the unperturbed data. 
In Figure \ref{fig:bf_cooks_dist}, this individual corresponds to the largest Cook's
distance in the unperturbed data. 
However, even with minimal perturbation, the maximum
Cook's distance becomes much lower, suggesting that this
individual is no longer as easily identifiable in the perturbed data.

 
\textit{Mahalanobis distance and minimum Euclidean distance:}

\begin{figure}[tb]
        \centerline{
         \includegraphics[width=5.0in]{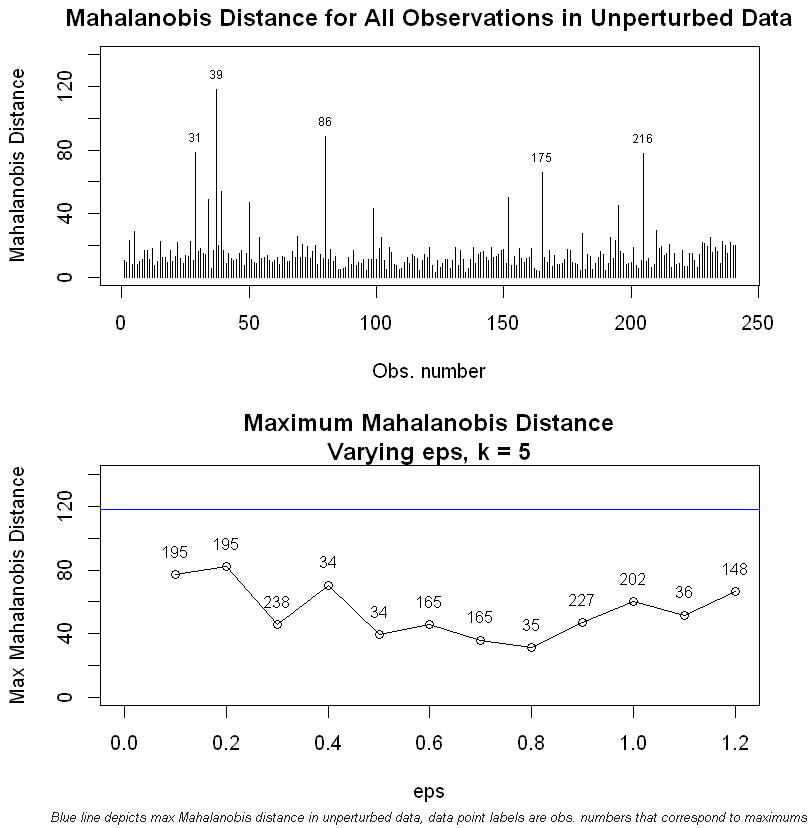}
        }
        \caption{Body Fat Data, Mahalanobis Distances}
        \label{fig:bf_mah_dist}
    \end{figure}

\begin{figure}[tb]
        \centerline{
         \includegraphics[width=5.0in]{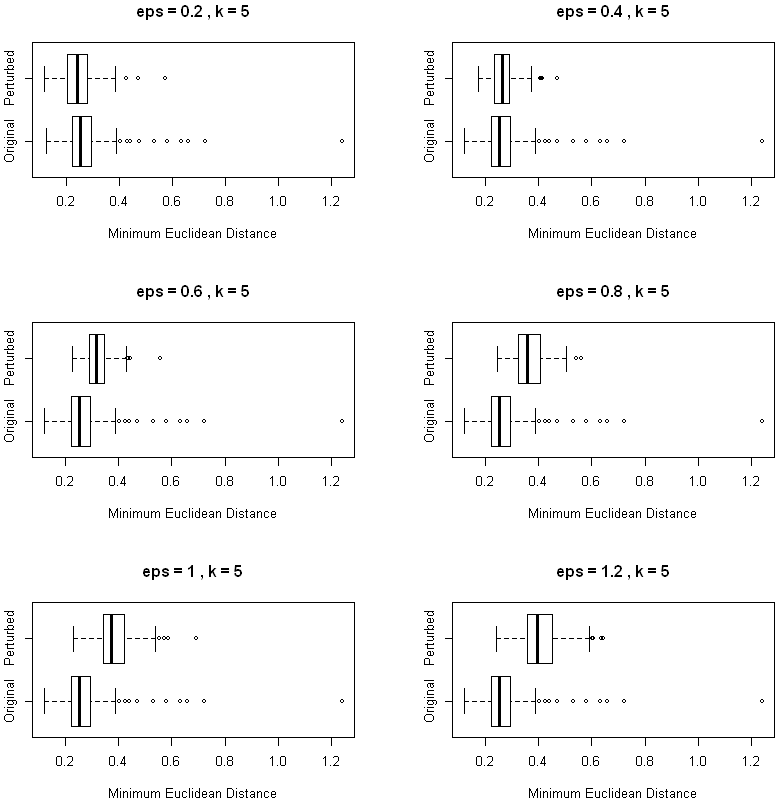}
        }
        \caption{Body Fat Data, Boxplots of Minimum Euclidean Distances
        Before and After Perturbation}
        \label{fig:bf_mindist_boxplots}
    \end{figure}

Figures \ref{fig:bf_mah_dist} and \ref{fig:bf_mindist_boxplots} 
show a similar result as the Cook's distance plots;
privacy is provided to identifiable data points for even small
values of $\epsilon$.


\textit{Principal component analysis:} 

After scaling the data, we performed principal component analysis on
both the unperturbed and perturbed datasets. For small values of
$\epsilon$, we find that the standard deviation of the first principal
component and proportion of variance corresponding to the first
principal component are slightly higher than in the unperturbed data. As
$\epsilon$ increases the variation appears to be spread over more
principal components. 


\begin{figure}[tb]
\centerline{
\includegraphics[width=5.0in]{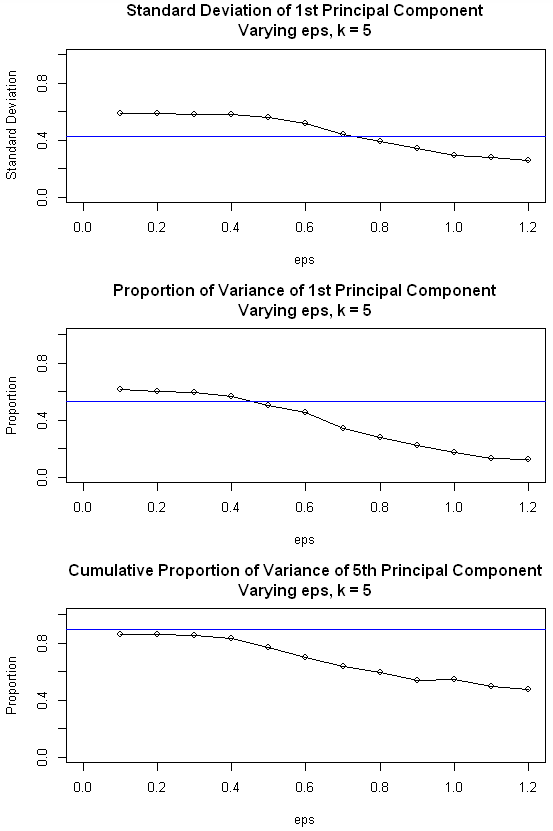}
}
\caption{Body Fat Data, PCA Variance Plots}
\label{fig:bf_pca_variance}
\end{figure}

%

\subsection{Prediction-Oriented Assessment}
\label{predict}

Statistical applications tend to fall into one of two general
categories, which we will refer to as Description and Prediction.  The
former has the goal of understanding some entity, process, effect and so
on, while the latter concerns predicting new data.

Much of the SDC literature has been aimed at the Description side of
things, estimating means, totals, regression coeficients and the like.
The Prediction side has rarely been the focus, and we turn to that
aspect in this section, particularly classification settings, the form
of many modern applications.

Intuitively, SDC methods should do fairly well in classification
settings.  Consider 2-class problems, for instance, where we are
predicting $Y = 0,1$ from a vector of covariates $X$, say with a
continuous distribution.  Assuming equal misclassification costs,
the set of $t$ for which $P(Y = 1 | X = t) = 0.5$ forms the decision
boundary.  What is the effect of data perturbation?

Consider a nonparametric regression method, say random forests.  If a
data point in the training set is far from the boundary, perturbation
will have little or no effect on future predictions; only points near
the boundary have much impact.  One may conjecture, then, that SDC
methods will not compromise prediction ability much, at least for
nonparametric regression methods.  This was confirmed in our
experiments.

We first looked at \textbf{pef}, a dataset included in the R
\textbf{regtools} package.  This is data on programmers and engineers,
from the 2000 US census.  We predict a variable \textbf{occ}, which
codes one of six occupations, from variables such as age, income and
education.  Here we took $q = 0.5$, for $k = 5, 10, 25, 50$.
Misclassification rates are as follows:

\begin{tabular}{|r|r|r|r|r|r|}
\hline
method & no SDC & k = 5 & k = 10 & k = 25 & k = 50 \\ \hline 
RF & 0.628 &0.627 & 0.645 & 0.682 & 0.659 \\ \hline
\end{tabular}

This dataset does not lend itself to strong predictability, with an
error rate about about 63\%.  However, that rate increases only slightly
under SDC.
The above results were based on 25 replications, with a holdout set size
of 1000.  

Here is the same analysis on a second dataset, the well-known Pima
diabetes study.  It's quite different from the census data, in that it
is much smaller (768 rows, vs. 20090 for \textbf{pef}), thus requiring
more privacy protection.  On the other hand, greater predictive accuracy
is possible for this data.

\begin{tabular}{|r|r|r|r|r|r|}
\hline
method & no SDC & k = 5 & k = 10 & k = 25 & k = 50 \\ \hline 
RF & 0.241 & 0.242 & 0.233 & 0.245 & 0.241 \\ \hline
\end{tabular}

Again, performance appears not to decline due to the privacy action, and
may even help, due to salutory smooting effects.

\section{Large Data Sets and Computational Complexity} 
\label{large}

Data sets are growing rapidly in both size and dimension.  This is driven by many factors, including
\begin{itemize}
  \item	A proliferation of data sources, including apps, back-office systems like ERP, smart devices, and sensors.
  \item	The growth of the systems themselves as businesses and other organizations achieve global scale.
  \item	Increases in data collection and storage capacity through networks and mass storage.
  \item	Larger computing devices and cloud scale computing.
  \item	The transition from recording data about entities (like people) to storing transactions like purchases to storing events like clicks.  This has happened as analysts go from trying to understand people to understanding what they do transactionally to how they behave and where they go.
  \item	The transition from structured to semi-structured to unstructured data.  Structured data, like that found in traditional database tables, is strictly constrained in terms of dimension.  Semi-structured data like JSON objects, which is often captured from running applications, is unconstrained in dimension.  Unstructured data, like documents, photos, or videos, is of almost unlimited dimension.  Unstructured data might not seem like a candidate for the method described in this paper, but documents, photos, and videos are often reduced to a set of features described by categorical, binary, or numeric variables.
\end{itemize}
As a result, datasets can now easily be in the billions of rows with hundreds or thousands of variables or features.  But with the growth of datasets comes a growth in risk.  Larger datasets put more people at risk, and increased dimensionality increases damage per person.  Also, the increase in dimensionality makes it easier to identify someone’s record in a dataset.
We need to be able to apply the method to these huge datasets, so the algorithm must be reasonably efficient from a computational complexity perspective.  As we will see in the next section, the basic algorithm is computationally expensive, but with slight modification can handle large datasets.

\section{Computational Complexity of the Basic Method}
To assess computational complexity of the method, going forward we will assume $p$ is large but fixed, $q=1$, and $n$ is increasing.  We will also assume that the distance metric used to define closeness is arbitrary.  This is reasonable, since we need wide latitude to define closeness in different ways for very different datasets.  Under these conditions, whether we are selecting nearby points based on epsilon neighborhood or $k$-nearest neighbor, the computational complexity of the basic method is $O(n^2)$, since we have to calculate the distance between all possible points.  This is probably not tenable for large datasets, e.g., $=10^9$, for which the number of calculations would be of the order $10^{18}$.

\section{Alternative Methods}

\subsection{Method 1:  Draw Neighbors from a Sample of Points}
For the first method, we will just take a smaller sample of data points to use as neighbors.  That is, we will take a sample S of size $m<<n$.  For each data point in the original data set, select the neighbors from the sample, then apply the method as before.  For this method, the complexity is $O(mn)$, considerably better than the original method.  For $n=10^9$, $m=10^4$, complexity is of order $10^{13}$ instead of $10^{18}$, an improvement of five orders of magnitude.
Assuming that all the data is in memory, the cost of generating a sample of size $m$ is $O(m)$, so if we were to use a distinct random sample for each point, the complexity would still be $O(mn)$. 
For $n=10^9$, $m=10^4$, complexity is still order $10^{13}$.  The advantage is that we potentially get slightly richer data and better protection.

\subsection{Method 2:  Sample Randomly from the Distance Matrix}
In Method 2, we are actually sampling twice as many points as we need to, since when we sample point $x_j$ to be a possible neighbor of the point $x_i$, we can also use $x_i$ as a potential neighbor of $x_j$.  This is due to the symmetry of distance metrics and the distance matrix
$D = [d_{ij}]$
where $d_{ij}$ is the distance between points $i$ and $j$.
So instead of drawing a new sample for each data point, we could simply draw a sample of elements from the entire distance matrix.  Then for each data point, we could look at the distances in the sample and find those that are sufficiently small or find the smallest $k$ elements.

We will now describe the method more formally and derive the complexity.  Let $W = (x_{ij})$ be the original data set.
We want to sample pairs $(x_i, x_j)$, $i<j$ at random with a sample size of
$n_s = n{m}/2$
where $m<<n$ is the desired sample size per data point.  There are $n(n-1)/2$ such pairs total, and we are sampling $n_s<< n(n-1)/2$.  To generate the sample, we can use a mixed congruential random number generator to generate numbers in $[1, n(n-1)/2]$.




Once we have $n_s$ random integers in the range $[1, n(n-1)/2],$ we will map them to $(i,j)$ pairs, $1<=i<j<=n$ to obtain the set $S_d = \{(i,j)\}$.  We then calculate the distances
$d_{ij}$ = $<xi,xj>$
and store them in an undirected graph $G_d$ with nodes representing the $n$ data points and edges
$\{e_{ij}: (i,j) \in S_d\}$
To map a random integer $r$ to an $(i,j)$ pair, we will use the following algorithm:
\begin{enumerate}
	\item Calculate $v = \sqrt{(8r+1)}$ as an integer operation.  That is, the calculation of $v$ produces an integer plus a remainder that indicates whether $v$ is the exact square root.
	\item If $v$ is odd and the exact square root,
	\\$j \leftarrow (1+v)/2$
	\\$i \leftarrow j - 1$
	\\and we’re done.
	\item Else
	\begin{enumerate}
		\item If $v$ is odd but not the exact root, \\$j \leftarrow (3+v)/2$
		\item Else ($v$ even)
		\\$j \leftarrow (2+v)$
		\\$i \leftarrow r – (j-1)(j-2)/2$
	\end{enumerate}
\end{enumerate}
Since the complexities of generating the random numbers, mapping the random numbers to $(i,j)$ pairs,
calculating distances, and finding neighbors are all $O(nm)$,
the complexity of the entire algorithm is $O(nm)$.

\subsection{Method 3:  Partitioning}
With the advent of cloud computing, it has become much more feasible to harness the power of many computers or virtual machines, each with gigabytes of memory.  Under this scenario, the problem (and data) are partitioned and spread across instances.  There are several ways this approach could be applied to the problem at hand:
\begin{enumerate}
	\item Simply partition the dataset into $u$ equal partitions and then apply the original method.  In this case, the size of each partition is $n/u$, so the complexity is $O(n^2/u^2)=O(n^2)/u^2$, a reduction by a factor of $u^2$ over the original method.
	\item Partition the dataset into u equal partitions and then apply the Method 2 above.  In this case, the size of each partition is $n/u$, so the complexity is $O(mn/u^2)=O(n^2)m/nu^2$, a reduction by a factor of $nu^2/m$ over the original method.
\end{enumerate}
Partitioning has two advantages:  It both spreads the work across multiple instances, thereby increasing the computing power that can be brought to bear on the problem, and it also reduces the complexity of the problem being solved on each node.  However, the resulting combined dataset will be somewhat different, since it is a combination of perturbed subsets.



\bigskip

\bibliographystyle{acm}
\bibliography{Paper}  

\end{document}